\pdfoutput=1
\documentclass[prl,twocolumn,preprintnumbers,nofootinbib,superscriptaddress,10pt]{revtex4-1}
\usepackage[dvips]{graphicx}

\usepackage{amsmath,amsthm,amssymb,multirow,psfrag}
\usepackage[normalem]{ulem}
\usepackage{epsfig}
\usepackage{color}
\usepackage{hyperref}
\graphicspath{{./Figures/}}

\begin{document}

\def\lsim{\mathrel{\rlap{\lower4pt\hbox{\hskip1pt$\sim$}}
  \raise1pt\hbox{$<$}}}
\def\gsim{\mathrel{\rlap{\lower4pt\hbox{\hskip1pt$\sim$}}
  \raise1pt\hbox{$>$}}}
\newcommand{\vev}[1]{ \left\langle {#1} \right\rangle }
\newcommand{\bra}[1]{ \langle {#1} | }
\newcommand{\ket}[1]{ | {#1} \rangle }
\newcommand{\ev}{ {\rm eV} }
\newcommand{\kev}{{\rm keV}}
\newcommand{\Mev}{{\rm MeV}}
\newcommand{\Gev}{{\rm GeV}}
\newcommand{\Tev}{{\rm TeV}}
\newcommand{\mw}{$M_{W}$}
\newcommand{\Ft}{F_{T}}
\newcommand{\Zparity}{\mathbb{Z}_2}
\newcommand{\BLambda}{\boldsymbol{\lambda}}
\newcommand{\met}{\;\not\!\!\!{E}_T}
\newcommand{\beq}{\begin{equation}}
\newcommand{\eeq}{\end{equation}}
\newcommand{\bea}{\begin{eqnarray}}
\newcommand{\eea}{\end{eqnarray}}
\newcommand{\nn}{\nonumber\\}
\newcommand{\gev}{{\mathrm GeV}}
\newcommand{\hc}{\mathrm{h.c.}}
\newcommand{\eps}{\epsilon}
\newcommand{\bwt}{\begin{widetext}}
\newcommand{\ewt}{\end{widetext}}
\newcommand{\draftnote}[1]{{\bf\color{blue} #1}}
\newcommand{\vecsigma}{ \mbox{\boldmath $\sigma$}}

\newcommand{\Mpl}{M_{\rm Pl}}
\newcommand{\TRH}{T_{\rm RH}}
\newcommand{\TCMB}{T_{\rm CMB}}
\newcommand{\xiend}{\xi_{\rm end}}
\newcommand{\Hend}{H_{\rm end}}
\newcommand{\HRH}{H_{\rm RH}}
\newcommand{\aRH}{a_{\rm RH}}
\newcommand{\aend}{a_{\rm end}}
\newcommand{\amax}{a_{\rm max}}
\newcommand{\abar}{a_{\rm bar}}
\newcommand{\kend}{k_{\rm end}}
\newcommand{\kmax}{k_{\rm max}}
\newcommand{\xend}{x_{\rm end}}
\newcommand{\kast}{k_\ast}
\newcommand{\rhoAend}{\rho^{\rm{end}}_E}
\newcommand{\rhoIend}{\rho^{\rm{end}}_I}
\newcommand{\sigend}{\overline{\sigma}_{\rm{end}}}	
\newcommand{\sigRH}{\overline{\sigma}_{\rm{RH}}}
\newcommand{\rhochiend}{\rho^{\rm{end}}_\chi}
\newcommand{\mbchi}{\overline{m}_\chi}
\newcommand{\sigb}{\overline{\sigma}}

\newcommand{\cO}{{\cal O}}
\newcommand{\cL}{{\cal L}}
\newcommand{\cM}{{\cal M}}

\newcommand{\fref}[1]{Fig.~\ref{fig:#1}} 
\newcommand{\eref}[1]{Eq.~\eqref{eq:#1}} 
\newcommand{\aref}[1]{Appendix~\ref{app:#1}}
\newcommand{\sref}[1]{Section~\ref{sec:#1}}
\newcommand{\tref}[1]{Table~\ref{tab:#1}}

\newcommand{\LU}[1]{\textcolor{red}{[LU: #1]}}



\title{\Large{{\bf Super heavy dark matter from inflationary Schwinger production}}}

\author{\bf {Mar Bastero-Gil}}
\email{mbg@ugr.es}
\affiliation{\normalsize\it
Departamento de F\'{i}sica Te\'{o}rica y del Cosmos and CAFPE,~Universidad de Granada, Campus de Fuentenueva, E-18071 Granada, Spain}

\author{\bf{Paulo B. Ferraz}}
\email{paulo.ferraz@student.uc.pt}
\affiliation{\normalsize\it
Departamento de F\'{i}sica Te\'{o}rica y del Cosmos and CAFPE,~Universidad de Granada, Campus de Fuentenueva, E-18071 Granada, Spain}
\affiliation{\normalsize\it
Universidad de Coimbra, Faculdade de Ciências e Tecnologia da Universidade
de Coimbra and CFisUC, Rua Larga, 3004-516 Coimbra, Portugal}

\author{\bf{Lorenzo Ubaldi}}
\email{lorenzo.ubaldi@ijs.si}
\affiliation{\normalsize\it
Jo\v{z}ef Stefan Institute, Jamova 39, 1000 Ljubljana, Slovenia}
\affiliation{\normalsize\it
Institute for Fundamental Physics of the Universe (IFPU), \\ Via Beirut 2, 34014 Trieste, Italy}

\author{\bf{Roberto Vega-Morales}}
\email{rvegamorales@ugr.es\\}
\affiliation{\normalsize\it
Departamento de F\'{i}sica Te\'{o}rica y del Cosmos and CAFPE,~Universidad de Granada, Campus de Fuentenueva, E-18071 Granada, Spain}

\preprint{UG-FT 328-23,~CAFPE 198-23,CA21106,~CA21106}
\begin{abstract}
We consider a simple setup with a dark sector containing dark electrons charged under an abelian $U(1)_D$ gauge symmetry.~We show that if the massless dark photon associated to the $U(1)_D$ is produced during inflation in such a way as to form a classical dark electric field, then dark electron-positron pairs are also produced close to the end of inflation via the Schwinger effect even if they are very massive.~For large enough dark electric force, dark electrons with masses larger than the Hubble scale can be produced which are non-relativistic at production and throughout their cosmic evolution.~They can account for the dark matter abundance today for masses in the range $\sim$\,100\,GeV to $10^{17}$\,GeV and up to six orders of magnitude larger than the Hubble scale at the end of inflation where purely gravitational production is exponentially suppressed.~We examine the regime where the dark electrons do not thermalize with the dark photons throughout their cosmic history and assume negligible kinetic mixing with the visible $U(1)$ so they remain decoupled from the Standard Model thermal bath as well.~Thus the final dark matter relic abundance is determined only by the initial inflationary Schwinger production and redshifting after reheating.

\end{abstract}
\maketitle

\section{Introduction} \label{sec:intro} 

The nature and production mechanism of dark matter is still unknown.~Significant research effort has been put into candidates like weakly interacting massive particles, axions, sterile neutrinos, but we have seen no experimental evidence for them yet~\cite{Bertone:2018xtm}.~This motivates studying alternative candidates as well.~One class which has received considerable attention in the recent past is that of dark photon dark matter, produced via non-thermal mechanisms~\cite{Graham:2015rva,Agrawal:2018vin,Dror:2018pdh,Co:2018lka,  Bastero-Gil:2018uel, Long:2019lwl,McDermott:2019lch,Nakai:2020cfw,Ahmed:2020fhc,Kolb:2020fwh,Salehian:2020asa,Bastero-Gil:2021wsf,Gorghetto:2022sue,Sato:2022jya,Redi:2022zkt,Barrie:2022mma} where the dark photon is associated with a dark $U(1)_D$ abelian gauge theory.~At some point before matter radiation equality the dark photon must become massive in order to provide a cold (non relativistic) dark matter candidate.~However, it has been argued~\cite{East:2022rsi} that the formation of vortices in these scenarios spoils the possibility of the dark photon being dark matter, though this can perhaps be avoided in some scenarios~\cite{Cyncynates:2023zwj}.

One can consider adding scalar and/or fermion fields charged under the $U(1)_D$ to the dark sector'\footnote{We will generically refer to both scalars and fermions charged under the dark  $U(1)_D$ as `dark electrons'.}, of which the dark photon is the gauge boson.~Such a `dark QED' setup was studied in~\cite{Arvanitaki:2021qlj} where the authors considered a massive dark photon whose longitudinal component is produced via inflationary fluctuations as in~\cite{Graham:2015rva} which is coupled to dark electrons.~They showed how via various dark QED processes, such as dark electromagnetic cascades and plasma dynamics, the dark electrons are produced and eventually thermalize with the dark photons while remaining decoupled from the visible thermal bath.~In the end it was found that the dark electrons can constitute the dark matter in a mass range between 50 MeV and 30 TeV.~An important point~\cite{Graham:2015rva, Arvanitaki:2021qlj} is that the dark photons produced during inflation form a `condensate' which can only be treated as a classical dark electric field after inflation in the radiation dominated era once their momentum modes re-enter the horizon.~It is only then that the processes leading to the production of dark electrons and to thermalization can begin taking place.

In this letter we consider a dark QED scenario as well, but utilize a mechanism where the classical dark electric field is present already by the end of inflation~\cite{Bastero-Gil:2018uel,Bastero-Gil:2021wsf}.~We show that in such a case it is possible to generate by the end of inflation a relic of dark electrons which today matches the observed dark matter abundance.~The dark photon can be massless during inflation as opposed to the scenario in~\cite{Graham:2015rva,Arvanitaki:2021qlj} where it is necessarily massive.~Furthermore, the dark electrons are produced entirely via the Schwinger effect during inflation as opposed to via thermalization within the dark sector after inflation as in~\cite{Graham:2015rva,Arvanitaki:2021qlj}.~Thus the final relic abundance is set by the initial inflationary produced Schwinger pairs followed by redshifting effects.~For large enough dark electric force, but small enough to not thermalize, dark electrons with masses larger than the Hubble scale can be produced which are non-relativistic throughout the entirety of their cosmic evolution.~They can account for the dark matter abundance today over a huge mass range from $\sim$\,100 GeV to as heavy as $\sim10^{15}$\,GeV for the entire allowed range of scales at the end of inflation $100$\,GeV $\lesssim \Hend \lesssim 10^{13}$\,GeV and a wide range of reheating temperatures.

\section{Dark Schwinger production}\label{sec:schwinger}

We start by introducing the dark electrons which we label $\chi$.~We parametrize the background de-Sitter spacetime by the Friedmann-Robertson-Walker metric with $ds^2 = - dt^2 + a^2(t) d\vec{x}^2$ and consider the action,
\bea\label{eq:dQEDaction}
S&=&
\int d^4x \sqrt{-g}
\left(
-\frac{1}{4} F^{\mu\nu}F_{\mu\nu} +
\, \mathcal{L}_{\rm{charged}}(A_\mu,\chi) 
\right) \, .~~
\eea 
For simplicity and for concreteness we focus on minimally coupled scalar dark electrons in this work, but our results also apply to fermions up to numerical factors of order one.~We thus consider the lagrangian,
\bea \label{eq:lagscal}
\mathcal{L}_{\rm charged}^{\rm scalar} = -|D_\mu \chi|^2 - m_\chi^2 |\chi |^2 \, ,
\eea
where we have the covariant derivative $D_\mu = \partial_\mu + i g_D A_\mu$, with $g_D$ the $U(1)_D$ dark gauge coupling. 

In the presence of a constant background dark electric field $E$ during inflation, such as the one discussed below (see~\eref{Efield}), pairs of dark electrons and positrons are produced via the Schwinger effect.~To see this we rely on previous studies of magnetogenesis~\cite{Kobayashi:2014zza,Hayashinaka:2016qqn,Banyeres:2018aax} which computed the Schwinger current $J$ of the generated $\chi$ charged particles.~To estimate the dark electron energy density $\rho_\chi$ we consider~\cite{Sobol:2019xls} the evolution equation,
\bea\label{eq:rhochievo}
\dot{\rho}_{\chi} + 3H\rho_{\chi} =  E \cdot J 
= \sigma  E^2  \, ,
\eea
where we have introduced the conductivity $\sigma = J/E$ and taken $\chi$ to be non relativistic during inflation since we are interested in producing super heavy dark matter with mass larger than the Hubble scale to be discussed more below.~This evolution equation can be derived from the energy momentum tensor and expresses energy conservation where $ E\cdot J $ acts as a source term for the $\chi$ energy density.~A corresponding term with the opposite sign appears in the evolution equation for the energy density of the dark electric field which is sourced by the inflaton, but depleted by the $- \sigma E^2$ term.~The latter is negligible when $\sigma/H \ll 1$ in which case the dark electric field is unaffected by backreaction from the Schwinger produced dark electrons.~This also implies a maximum dark gauge coupling such that these backreaction effects can be neglected and which we limit ourselves to in this study.

As is well known, gravitational production of particles with masses larger than the Hubble scale is exponentially suppressed by $m_\chi/H$.~To overcome this suppression requires a strong dark electric force where $g_D E > H^2$.~In this regime dark Schwinger production dominates over purely gravitational effects.~The conductivity can be computed in terms of two dimensionless ratios~\cite{Kobayashi:2014zza,Hayashinaka:2016qqn,Banyeres:2018aax},
\bea\label{eq:ratios}
\lambda \equiv \frac{ g_D E }{ H^2 } , \quad \mbchi \equiv \frac{ m_\chi }{ H} ,
\eea
with which one can study different limits.~The strong electric force regime is defined via the limit $\lambda \gg 1$.~In this regime the Schwinger effect exhibits the same behaviour in de-Sitter space as in flat space and leads to the dimensionless conductivity~\cite{Kobayashi:2014zza,Banyeres:2018aax},
\beq \label{eq:sigbarstrong}
\bar\sigma \equiv \frac{\sigma}{H} = \frac{g_D^2}{12\pi^3} \lambda e^{- \pi \mbchi^2/\lambda}
~~(\lambda \gg 1) .
\eeq
This applies to scalars, while for fermions the conductivity is larger by a factor of 2~\cite{Hayashinaka:2016qqn}.~When $\lambda \gg \mbchi$ we see the conductivity is not exponentially suppressed even if $m_\chi > H$.~This implies that particle production can occur during inflation via the Schwinger effect even if purely gravitational effects are negligible.~As we'll see, there is a sizable parameter space in which inflationary Schwinger production can generate the observed dark matter relic abundance for dark electron masses with $m_\chi > \Hend$.

As we consider the Schwinger production during inflation, it is convenient to rewrite \eref{rhochievo} switching to conformal time, $d\tau = dt/a$, and use $\tau = -1/(aH)$ to obtain for the evolution equation,
\bea
\partial_{\tau}\rho_{\chi}-\frac{3}{\tau}\rho_{\chi} = 
- \bar{\sigma}E^2\frac{1}{\tau} .
\eea
Evolving from $\tau_0$, taking $\rho_\chi(\tau = \tau_0) = 0$ as the initial condition, to the end of inflation $\tau_{\rm end}$ one finds,
\bea\label{eq:rhochi}
\rho_\chi^{\rm end} \equiv \rho_\chi(\tau_{\rm end})=  \frac{\bar{\sigma}}{3}E^2
 \Big[1-\Big(\frac{\tau_0}{\tau_{\rm end}}\Big)^{-3}\Big] \simeq \frac{\bar\sigma_{\rm end}}{3}E^2 \, .~~
\eea
We have used $\tau_0/\tau_{\rm end} = e^{N_e}$, with $N_e$ the number of e-folds between $\tau_0$ and $\tau_{\rm end}$, to reach the last approximate equality.~Even if we have a dark electric field present only for a few e-folds close to the end of inflation, the solution in~\eref{rhochi} holds. Since $\bar\sigma \ll 1$, we also have that $\rho_\chi^{\rm end}$ is much smaller than the energy density of the electric field which justifies neglecting the backreaction of $\chi$ on $E$ thus potentially spoiling the Schwinger effect.

\section{Generating a dark electric field} \label{sec:electric}

We consider an abelian $U(1)_D$ group whose gauge boson is the dark photon.~A simple way of generating a classical dark electric field during inflation is by considering the coupling $\frac{\alpha}{4f} \phi F_{\mu\nu} \tilde F^{\mu\nu}$ where $\phi$ is the inflaton, $F_{\mu\nu}$ is the dark photon field strength~\cite{Bastero-Gil:2018uel,Bastero-Gil:2021wsf}.~The coupling leads to the exponential production of one of the two transverse polarizations of the dark photon.~The typical wavelength of the produced dark photons is roughly the size of the Hubble horizon.~This produces a classical dark electric field~\cite{Anber:2009ua,Bastero-Gil:2021wsf} that at the end of inflation has magnitude approximately given by,
\beq \label{eq:Efield}
E \approx 10^{-2} \frac{e^{\pi \xiend}}{\xiend^{3/2}} \Hend^2 \, ,
\eeq
which is coherent within the horizon with energy density,
\beq \label{eq:rhoE}
\rho_E = \frac{1}{2}E^2 .
\eeq
Here $\xi \equiv \frac{\alpha \dot\phi}{2Hf}$ with $H$ the Hubble parameter.~The subscript `end' indicates that the quantities are evaluated at the end of inflation.~We consider values of $\xiend \lesssim 9$ in which case we can neglect the backreaction of the produced dark electric field on the inflaton dynamics~\cite{Bastero-Gil:2021wsf}.~An accompanying dark magnetic field is also produced, but it is suppressed by $\rho_B = \rho_E/\xi^2$.~It is also automatically produced parallel to the electric field~\cite{Bastero-Gil:2023mxm} and would thus lead to a small enhancement of the Schwinger effect~\cite{Fujita:2022fwc}, but for the rough estimates of the parameter space we make here it can be neglected.~In this mechanism for generating the dark electric field, the dark photon can be massless during inflation as the production involves only the transverse polarizations.~This is in contrast to~\cite{Graham:2015rva, Arvanitaki:2021qlj} where the dark photon must be massive during inflation in order to produce the longitudinal mode.

\section{Cosmic evolution}\label{sec:evolution}

For simplicity, we approximate the reheating process that takes us from the end of inflation to the radiation dominated era as instantaneous.~At the end of inflation the inflaton dominates the energy density of the Universe,
\bea \label{eq:rhoIend}
\rhoIend = 3 \Hend^2 \Mpl^2 .
\eea
It must transfer some of this energy density into visible radiation whose energy density we parametrize as,
\bea \label{eq:rhoRH}
\rho_R (T_{\rm RH})  
= \frac{\pi^2}{30}g_*(\TRH) \TRH^4 \, 
\equiv \epsilon_R^4 \rhoIend 
\eea
where $\epsilon_R<1$ parametrizes the efficiency of reheating and $g_*(\TRH)$ is the number of relativistic degrees of freedom which we fix to $g_*(\TRH) \sim 100$ and restrict ourselves to reheating temperatures above the electroweak scale.~Given the inflaton energy density at the end of inflation in~\eref{rhoIend}\, we see~\eref{rhoRH} also defines the reheating temperature $\TRH$.~At reheating we take the energy densities of dark electric field and dark electrons as equal to those at the end of inflation, $\rho_E^{\rm RH} = \frac{1}{2}E^2$ and $\rho_\chi^{\rm RH} =  \frac{\bar\sigma_{\rm end}}{3}E^2$ with $E$ given by \eref{Efield}.~So we have the hierarchy $\rho_R (T_{\rm RH}) \gg \rho_E^{\rm RH} \gg \rho_\chi^{\rm RH} $. 

The expression for the conductivity in~\eref{sigbarstrong} was derived in~\cite{Kobayashi:2014zza} assuming a de-Sitter background so it does not hold 
in the epoch of radiation domination (RD) for which, as far as we know, there are no studies of the conductivity.~However, we presume it will maintain the same exponential suppression during RD of \,$ {\rm Exp}\left[ - \pi m_\chi^2/g_D E  \right] $.~After reheating the dark electric field (energy density) redshifts like radiation so the exponential suppression quickly becomes more severe.~We thus expect Schwinger pair production to be negligible during RD.~In this case the energy density of $\chi$ simply redshifts like non-relativistic matter from reheating until today,
\beq \label{eq:rhochinon}
\rho_\chi(T) = \frac{\bar\sigma_{\rm end}}{3}  E^2 \left(\frac{T}{\TRH} \right)^3 \, ,
\eeq
where the energy density today is found by setting $T = T_0 \approx 10^{-13}$\,GeV to today's CMB temperature.~The energy density in the electric field $\rho_E$, which redshifts like $T^4$, remains negligible compared to $\rho_R$ throughout the history of the universe and is much smaller than $\rho_\chi$ today.~One may worry that after reheating the electric field could accelerate the dark electrons effecting how their energy density redshifts.~However, after reheating the electric field quickly begins oscillating~\cite{Bastero-Gil:2021wsf,Bastero-Gil:2023mxm} and no longer efficiently accelerates the dark electrons so the physical momentum simply redshifts with expansion.

\section{Dark matter parameter space}\label{sec:pspace}

\begin{figure*}[t!]
\begin{center}
\includegraphics[width=0.3275\textwidth]{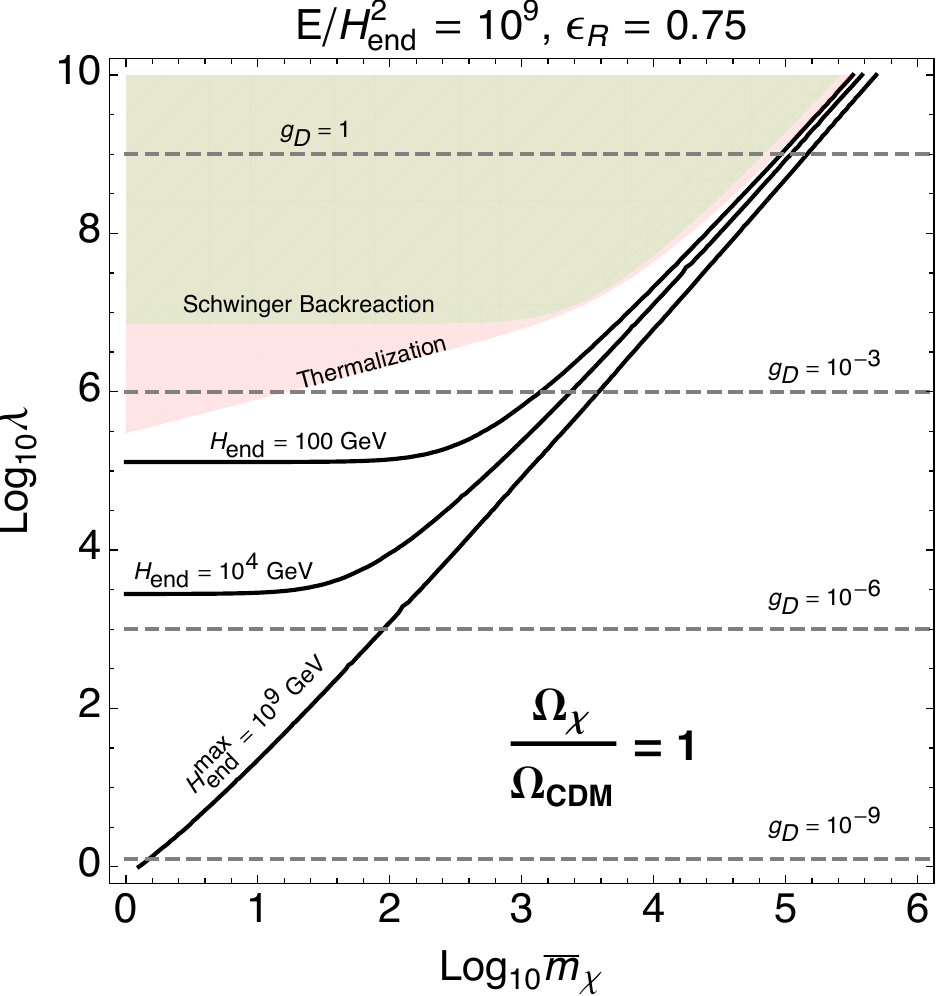}
\includegraphics[width=0.32\textwidth]{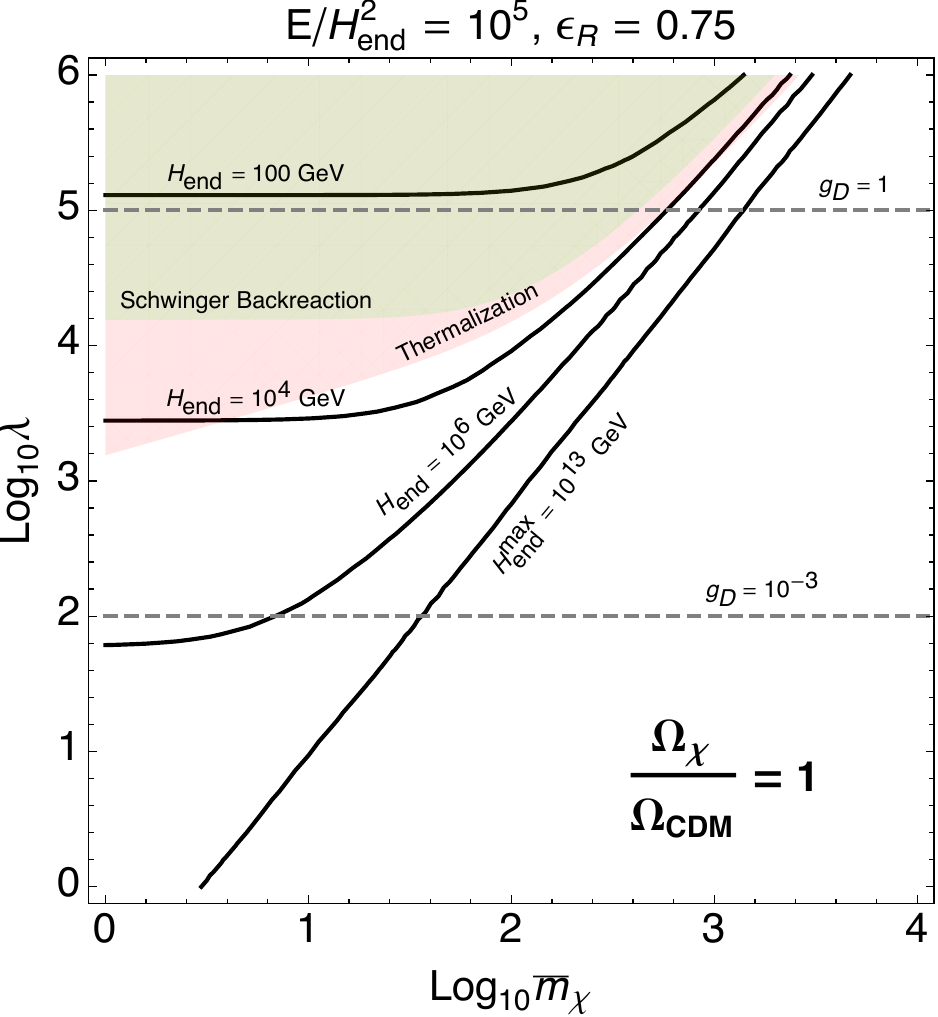}
\includegraphics[width=0.32\textwidth]{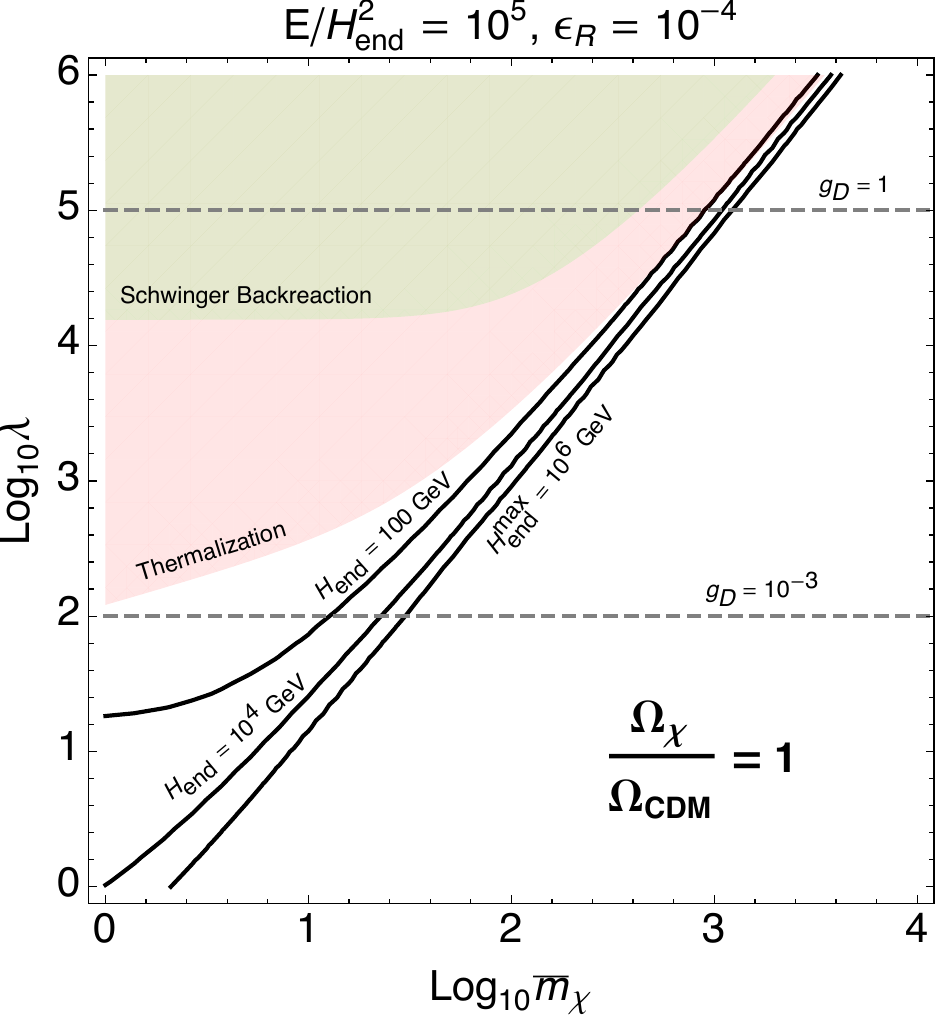}
\end{center}
\caption{Contours of $\Omega_\chi / \Omega_{\rm CDM} = 1$ (see~\eref{relic}) as a function of $\lambda$\,vs.\,$\mbchi$, where $\lambda \equiv g_D E/\Hend^2$ and $\mbchi \equiv m_\chi/\Hend$, for values of $E/\Hend^2$ and $\epsilon_R$  as indicated on top of each plot.~The Hubble parameter at the end of inflation is labeled $\Hend$ while $\epsilon_R$ parametrizes the fraction of the energy density transferred from the inflaton to the visible radiation at reheating (see text).}
\label{fig:ParSpace}
\end{figure*}

The above analysis assumes the various dark sector components redshift independently so we need to ensure that the dark sector does not thermalize within itself.~This requires that the rate of interactions between the dark charged particles and the dark photon is always slower than Hubble which we can estimate as follows.~Since $\chi$ is always non-relativistic, the cross section for $2 \leftrightarrow 2$ processes in the dark sector scales as
$\sigma_{2\leftrightarrow 2} \sim g_D^4 / m_\chi^2$.~The number density is $n_\chi(T) \simeq \rho_\chi(T)/m_\chi$ with $\rho_\chi(T)$ given in \eref{rhochinon}.~The interaction rate is $\Gamma(T) = n_\chi(T) \sigma_{2\leftrightarrow 2}$ and inherits the exponential suppression of the conductivity $\bar\sigma_{\rm end}$.~In the strong electric force limit and requiring $\Gamma(T) / H(T) \ll 1$ at reheating so there is no thermalization leads to the upper bound on $\lambda$,
\bea\label{eq:lambound1}
\lambda^7 e^{- \pi \mbchi^2/\lambda} \lesssim 12\pi^3\,
\epsilon_R^{2} \mbchi^{3} (\frac{E}{\Hend^2})^{4} .
\eea
There is an additional upper bound on $\lambda$ coming from requiring no backreaction on the electric field from the induced Schwinger current implying $\sigb < 1$ and leads to,
\bea\label{eq:lambound2}
\lambda^3 e^{- \pi \mbchi^2/\lambda} \lesssim 12\pi^3 \, (\frac{E}{\Hend^2})^{2} .
\eea
In general this is weaker than the no thermalization bound, but can be relevant for $\epsilon_R \sim 1$ (see~\fref{ParSpace}).

For the mechanism considered here, the dark vector could in principle be massless.~However, if the dark vector is massless limits from structure formation and plasma instabilities~\cite{Lasenby:2020rlf}, which are stronger than those coming from triaxiality of dark matter haloes~\cite{Lasenby:2020rlf,Agrawal:2016quu}, put severe constraints on the dark gauge coupling.~These constraints would rule out much of the dark matter parameter space for inflationary Schwinger production.~They can be avoided if the dark vector has a mass larger than $m_A \gtrsim 10^{-10}\,eV$ so at some point during its cosmic evolution, it must obtain a mass.~For masses of this order the dark photon contributes negligibly to the dark matter relic abundance and can be neglected when examining the dark matter parameter space below.

As discussed, we must also require the hierarchy in energy densities at reheating $\rhoIend \gg \rho_R (T_{\rm RH}) \gg \rho_E^{\rm RH} \gg \rho_\chi^{\rm RH} $.~Imposing that the visible radiation has energy density lower than the inflaton at the end of inflation ($\rho_R \ll \rhoIend$) imposes $\epsilon_R < 1$.~To ensure the Universe is radiation dominated until matter radiation equality requires that $\rho_R > \rhoAend$ and translates into an upper bound on
$\Hend$,
\bea\label{eq:Hendbound}
\Hend < \frac{ \sqrt{6} \, \epsilon_R^2  M_{\rm Pl} }{(E/\Hend^2)}  \, .
\eea
So we see that for very large electric fields or reheating temperatures which are low compared to the Hubble scale at the end of inflation $(\epsilon_R \ll 1)$, the upper bound on $\Hend$ becomes very stringent.~We impose the stronger between this upper bound and the absolute astrophysical upper bound of $\Hend \lesssim 10^{13}$\,GeV which must be smaller than the inflationary scale early during inflation.~We must also ensure no backreaction on the inflaton dynamics from production of the dark electric field.~This requires $\rhoAend < \rhoIend$ which restricts the size of the electric field to~\cite{Bastero-Gil:2018uel,Bastero-Gil:2021wsf},
\bea\label{eq:ratiobound}
1 < \frac{E}{\Hend^2} < 10^9 .
\eea
Note that for the dark photon production mechanism discussed above, this corresponds to $1.6 \lesssim \xiend \lesssim 9$.

As discussed, once the dark electrons are produced at the end of inflation they simply redshift like non-relativistic matter until today.~Thus to see if they can account for the observed dark matter relic abundance we only have to use~\eref{rhochinon} and set $\rho_\chi(T_0) = \rho_{\rm CDM} = 9.6 \cdot 10^{-48} \ {\rm GeV}^4$ with $T = T_0 \simeq 2.3 \cdot 10^{-13}$ GeV the current temperature of the universe.~We can then write the ratio $\rho_\chi(T_0) / \rho_{\rm CDM} = \Omega_\chi / \Omega_{\rm CDM}$ as,
\beq \label{eq:relic}
\frac{\Omega_\chi}{\Omega_{\rm CDM}} \simeq  
10^{12} \cdot \frac{\lambda^3 e^{-\pi \mbchi^2/\lambda} }{\epsilon_R^3} 
\left( \frac{\Hend}{10^{13} \ {\rm GeV} }\right)^{5/2} .
\eeq
where the dimensionless ratios $\lambda$ and $\mbchi$ are defined in~\eref{ratios}.~In~\fref{ParSpace} we present contours of $\Omega_\chi / \Omega_{\rm CDM} = 1$ as a function of $\lambda$\,vs.\,$\mbchi$ for values of $E/\Hend^2$ and $\epsilon_R$ as indicated on the top of each plot.~Contours for different values of $\Hend$ are shown where in order to satisfy $\rhoAend < \rhoIend$ the maximum value of $\Hend$ for a given $E/\Hend^2$ is indicated.~We also indicate the exclusion regions due to the no thermalization condition (red shaded) defined in~\eref{lambound1} and the no Schwinger backreaction condition (green shaded) defined in~\eref{lambound2}.~As can be seen, these conditions minimally constrain the parameter space so we take $\lambda$ to be bounded from above by requiring the dark gauge coupling remain perturbative.

We see for large enough dark electric fields with $E/\Hend^2 \sim 10^9$ and $\epsilon_R = 0.75$ (left) corresponding to very efficient reheating, as well as a dark gauge coupling near its perturbative limit ($g_D \sim 10$), ratios as large as $m_\chi/\Hend \sim 10^6$ can be obtained for a maximum Hubble scale at the end of inflation of $\Hend \sim 10^9$\,GeV.~Note that ratios of $m_\chi/\Hend \sim 10^6$ would be exponentially suppressed in purely gravitational production.~For moderately large electric fields with $E/\Hend^2 \sim 10^5$ (middle) ratios as large as $m_\chi/\Hend \sim 10^4$ can be obtained.~Since the maximum allowed Hubble scale is now $\Hend \sim 10^{13}$\,GeV, this allows for larger dark matter masses up to $m_\chi \sim 10^{17}$\,GeV.~When reheating is less efficient corresponding to $\epsilon_R = 10^{-4}$ (right) we are restricted to lower $\Hend \lesssim 10^6$\,GeV in order to ensure $\rhochiend < \rho_R$.~We see ratios as large as $m_\chi/\Hend \sim 10^4$ and dark matter masses as large as $m_\chi \sim 10^{10}$\,GeV can still be obtained in this case.

In summary the final dark matter parameter space available spans many orders of magnitude from $m_\chi \sim 100$\,GeV to $m_\chi  \sim10^{17}$\,GeV over the entire allowed range of Hubbles scales $100$\,GeV $\lesssim \Hend \lesssim 10^{13}$\,GeV.~These results hold generically for any mechanism which can generate the necessary background dark electric field.

\section{Summary and Discussion} \label{sec:sum} 

We have considered a minimal setup to explain the origin of dark matter consisting of a dark $U(1)_D$ gauge group containing a dark photon and dark electrons (scalar or fermion), $\chi$.~If the dark photons form a constant classical electric field during inflation, dark electrons are inevitably produced via the Schwinger effect even if they have a mass larger than the Hubble scale where gravitational production is exponentially suppressed.~The production occurs in the `strong electric force' regime where $g_D E > \Hend^2$.~Since the dark electrons have mass larger than the Hubble scale at the end of inflation, they are non-relativistic at production.~Their energy density is suppressed compared to the dark electric field which in turn is small compared to that stored in the inflaton (and radiation).~However, because the dark electrons redshift as non-relativistic matter after inflation, eventually they come to dominate the energy budget of the universe and can provide a viable cold dark matter candidate.~The observed dark matter relic abundance can be obtained for dark electron masses ranging from from $m_\chi \sim 100$\,GeV to as large as $m_\chi \sim10^{17}$\,GeV over the entire allowed range of Hubbles scales at the end of inflation $100$\,GeV $\lesssim \Hend \lesssim 10^{13}$\,GeV and a wide range of reheating temperatures.

The dark matter production mechanism discussed here is non-thermal.~The dark sector does not thermalize with itself and remains decoupled from 
the thermal bath of the visible sector as we assumed no coupling between the two sectors.~One could contemplate introducing a kinetic mixing between the dark and visible (electromagnetic) $U(1)$'s.~Since the dark electrons are in general heavy one could in principle allow for a large kinetic mixing and still avoid thermalization as the cross sections and reaction rates are suppressed.~A large kinetic mixing could give some leverage for possible experimental searches of this type of dark matter.~As discussed above, there could also be interesting implications for structure formation due to the dark coulomb interaction, but we leave a dedicated study of these interesting possibilities for future work.


~\\
~\\

\noindent
{\bf Acknowledgments:}~The authors thank Antonio Torres Manso, Manel Masip, Jose Santiago, Laura Lopez Honorez, Dipan Sengupta, Nicholas Orlofsky, and Takeshi Kobayashi for useful comments and discussions.~This work has been partially supported by Junta de Andaluc\'a Project A-FQM-472-
UGR20 (fondos FEDER) and by MICIU/AEI/10.13039/501100011033 and FEDER/UE (grant PID2022-139466NB-C21) (R.V.M.) as well as by MICINN (PID2019-105943GB-I00/AEI/10.13039/501100011033,PID2022.140831NB.I00
/AEI/10.13039/501100011033/FEDER,UE)\,(M.B.G.),\,
FCT-CERN\,grant\,No.\,CERN/FIS-PAR/0027/2021 (M.B.G., P.F.),\,FCT\,Grant\,No.SFRH/BD/151475/2021 (P.F.).~The work of LU was supported by the Slovenian Research Agency under the research core funding No. P1-0035 and in part by the research grant J1-4389.~This article is based upon work from COST Action COSMIC WISPers CA21106, supported by COST (European Cooperation in Science and
Technology).~RVM is grateful to the Mainz Institute for Theoretical Physics (MITP) of the Cluster of Excellence PRISMA$^+$ (Project ID 390831469), for its hospitality and partial support during the completion of this work.

\bibliographystyle{apsrev}
\bibliography{DarkSchwingerRefs}

\end{document}